\documentclass[preprint,showpacs,preprintnumbers,amsmath,amssymb,showkeys]{revtex4}
\usepackage{graphicx}% Include figure files
\usepackage{dcolumn}% Align table columns on decimal point
\usepackage{bm}% bold math

\begin{document}
\renewcommand{\thesection}{\arabic{section}}
\renewcommand{\thetable}{\Roman{table}}
\setlength{\baselineskip}{16.0pt}
\makeatletter
\renewcommand{\fnum@table}[1]{Table~\thetable. } % for tables if req.
\renewcommand{\fnum@figure}[1]{FIG~\thefigure. }
\makeatother

\bibliographystyle{apsrev}

\title{Excitation of surface dipole and solenoidal modes \\ on toroidal structures}

\author{M. Encinosa and M. Jack}
\affiliation{ Florida A\&M University Department of Physics \\
205 Jones Hall \\ Tallahassee FL 32307}
\email{encinosa@cepast.famu.edu}
 %\altaffiliation[Also at ]{Physics Department, XYZ University.}
%\author{M. Jack}%
 %\email{Second.Author@institution.edu}
%\affiliation{Florida A\&M University Department of Physics \\
%Tallahassee FL 32307}
%\affiliation{ DRAFT}

%\date{\today}% It is always \today, today,
             %  but any date may be explicitly specified
\begin{abstract}

The time dependent Schr\"odinger equation inclusive of curvature
effects is developed for a spinless electron constrained to motion
on a toroidal surface and subjected to circularly polarized and
linearly polarized waves in the microwave regime. A basis set
expansion is used to determine the character of the surface
currents as the system is driven at a particular resonance
frequency. Surface current densities and magnetic moments
corresponding to those currents are calculated. It is shown that
the currents can yield magnetic moments large not only along the
toroidal symmetry axis, but along directions tangential and normal
to the toroidal surface as well.

\end{abstract}

% PACS, the Physics and Astronomy
                             % Classification Scheme.
%Use showkeys class option if keyword
                              %display desired
\maketitle \setlength{\baselineskip}{18.0pt}
\section{Introduction}

The control of a nanostructure's state through electromagnetic
interactions is of fundamental and practical interest \cite{shapiro,
borzi, potz,qin, oosterkamp}. Considerable effort has been directed
towards the study of flat quantum rings, which, because of their
topology, give rise to Aharanov-Bohm and persistent current effects
\cite{bulaev2,climente,datta,filikhin, fuhrer,georgiev,
gylfad,ivanov,latil,pershin1,pershinring,
sasaki1,sasaki2,simonin,viefers}. Currents generated on a ring are
restricted to yield magnetic moments perpendicular to the plane of
the ring; however, it is conceivable that structures with different
topologies are capable of producing magnetic moments in different
directions, a feature potentially useful when such structures are
employed as nano-device elements. In this work, a toroidal structure
in the presence of an electromagnetic wave on the microwave scale is
considered. In contrast to rings, it is possible to excite modes for
which currents circulate around the minor radius of the toroidal
surface $T^2$ (called in what follows $ \lq\lq$solenoidal modes"
since the focus will be on magnetic moments) in addition to the
modes in which currents circulate azimuthally around the major
radius ($ \lq\lq$dipole modes").

The approach adopted here will be to develop
\begin{equation}
H = H_0 + V(t)
\end{equation}
with $H_0$ the Hamiltonian for a spinless electron constrained to
motion on $T^2$ inclusive of surface curvature effects
\cite{encmott}, and
 $V(t)$ the time dependent electromagnetic interaction with the
 form
appropriate to  circularly and linearly polarized microwaves
propagating along the symmetry axis of the torus (here the z-axis)
at a particular resonant frequency of the system. It has been shown
that for the case of no external vector potential, it is possible to
treat the normal degree of freedom as a spectator variable
\cite{encscripta}. However, inclusion of a vector potential
$\mathbf{A}$ with components normal to the surface complicates the
situation. Reducing the full three-dimensional problem to a
two-dimensional effective model serves to simplify matters
considerably. The resulting Schr\"odinger equation is then solved with
standard methods by taking ($\hbar$ = 1)
\begin{equation}
\Psi(t)= \sum_{n\nu}d_{n\nu}(t)\chi_n(\theta) {\rm exp}[i \nu
\phi]{\rm exp}[-i E_{n\nu }t]
\end{equation}
with the $\chi_n(\theta){\rm exp}[-i\nu \phi]{\rm exp}[-i E_{n\nu
}t]$ eigenstates of $H_0$. Surface current densities (SCDs) are
determined and
 magnetic moments are calculated,
from which it is shown that is it possible to selectively generate
dominant oscillatory magnetic moments in different directions
depending on the character of the applied signal.

The remainder of this paper is organized as follows: in section 2
the Hamiltonian on $T^2$ is developed by beginning with a three
dimensional formulation and proceeding to restrict the particle
(accounting for curvature)to $T^2$. In section 3 the eigenstates
of $H_0$ that comprise the basis set are given and the solution
method detailed. Section 4 presents results in the form of plots
of time dependent SCDs and magnetic moments. Section 5 is reserved
for conclusions.

\section{The system Hamiltonian}

This section presents a  derivation of $H(t)$ inclusive of surface
curvature (SC) effects. The necessity of accounting for SC has been
shown elsewhere \cite{encscripta}.  The central idea is upon
restricting a particle in the neighborhood of a  surface to the
surface proper, a geometric potential $V_C$ is necessary to capture
the full three dimensional spectra and wave functions with good
fidelity.

In this work, $V_C$ will not be the only geometric potential. The
normal component of $\mathbf A$ also couples to SC
\cite{pershinring,encinosa5}. It should also be noted that in
general, the Laplacian on the surface must be modified due to
curvature. In the case of $T^2$ the symmetry of the system precludes
such modifications \cite{encmott}.

The Schr\"odinger equation for a spinless electron in the presence
of a vector potential ${\mathbf A}(t)$ (reinserting $\hbar$) is
\begin{equation}
 H = {1 \over {2m}}\bigg ( {\hbar
\over i} \nabla + e {\mathbf A}(t) \bigg) ^2\Psi = i \hbar {
\partial \Psi \over \partial t}.
\end{equation}
To derive the gradient appearing in Eq. (3), let ${\mathbf
e}_{\rho},{\mathbf e}_{\phi},{\mathbf e}_{z}$ be unit vectors in a
cylindrical coordinate system. Points in the neighborhood of a
   toroidal surface of major radius $R$ and minor radius
 $a$  may then be parameterized by
\begin{equation}
 \mathbf{x} (\theta,\phi,q)= (R + a \ {\rm cos} \theta ){\mathbf
e}_{\rho} +a\  {\rm sin} \theta {\mathbf e}_{z} + q {\mathbf
e}_{n}
\end{equation}
with ${\mathbf e}_{n}$(to be given momentarily) everywhere normal
to $T^2$,  and $q$ the coordinate measuring the distance from
$T^2$ along ${\mathbf e}_{n}$. The  differential line element $d
\mathbf{x}$ is then
\begin{equation}
 d \mathbf{x}=a(1+k_1q){\mathbf e}_\theta  d\theta
+W(\theta,q)(1+k_2q){\mathbf e}_\phi d\phi + {\mathbf e}_n dq
\end{equation}
with
\begin{equation}
{\mathbf e}_{n} = \rm cos \theta  {\mathbf e}_{\rho}+\rm sin\theta
{\mathbf e}_{z}
\end{equation}
\begin{equation}
{\mathbf e}_{\theta} =-\rm sin \theta  {\mathbf e}_{\rho}+\rm cos
\theta {\mathbf e}_{z}
\end{equation}
\begin{equation}
W(\theta,q)= R + (a+q) \rm cos\theta
\end{equation}
and the toroidal principle curvatures
\begin{equation}
k_1 = {1 \over a}
\end{equation}
\begin{equation}
k_2 = {{\rm cos}\theta \over W(\theta,0)}.
\end{equation}
The result for $\nabla$ is then
\begin{equation}
\nabla = {\mathbf e}_{\theta} {1 \over (a+q)} {\partial \over
\partial \theta}+ {\mathbf e}_{\phi} {1 \over W(\theta,q)} {\partial \over \partial
\phi}+ {\mathbf e}_{n}  {\partial \over \partial q}.
\end{equation}
\vskip 6pt

To proceed with the development of $H(t)$  specific choices must
be made for ${\mathbf A}(t)$. Here the vector potential
corresponding to a CPW
\begin{equation}
{\mathbf A}(t) = {E_0 \over \omega}\big[-{\rm sin} (kz-\omega t)
{\mathbf i}+{\rm cos} (kz-\omega t) {\mathbf j}\big]
\end{equation}
with $E_0$ the field amplitude, $\omega$ the frequency and $z =
(a+q)\rm sin\theta$ is taken. The LPW case can be obtained
trivially from Eq. (12).

 The procedure by which $V_C$ is obtained from Eqs. (3)-(11) is well
 known (more detailed derivations can be found in
 \cite{burgsjens,jenskoppe,dacosta1,dacosta2,matsutani,goldjaffe,
exnerseba,schujaff}) and is summarized briefly below.

Begin by making a product ansatz for the surface and normal parts
of the wave function

\begin{equation} \Psi(\theta,\phi,q) \rightarrow
{\chi_S(\theta,\phi)\chi_N(q) \over (1 + 2q h(\theta)+ q^2
k(\theta))^{1 \over 2}}
\end{equation}
with $h(\theta), k(\theta)$ the mean and Gaussian curvatures of the
surface \cite{diffgeom}, and imposing conservation of the norm
through
\begin{equation}
|\Psi(\theta,\phi,q)|^2 (1+2qh(\theta)+q^2k(\theta)) d\Sigma dq
\rightarrow |\chi_S(\theta,\phi)|^2|\chi_N(q)|^2 d\Sigma dq
\end{equation}
 with $d\Sigma$ the differential surface area. The Laplacian
appearing in Eq. (3) takes the form (the subscript on the gradient
operator indicates only surface terms are involved)
\begin{equation}
\nabla^2 = \nabla_S^2 + 2h(\theta) {\partial \over \partial q} +
{\partial^2 \over
\partial q^2}
\end{equation}
with $q$ appearing in $\nabla_S^2$ as a parameter that can be
immediately set to zero. The differentiations in Eq. (15) act upon
the wave function given in Eq. (13) to yield
\begin{equation}
2h {\partial \over \partial q}+{\partial^2 \over \partial q^2}
\rightarrow {\partial^2 \over \partial q^2} +
h^2(\theta)-k(\theta)
\end{equation}
in the $q \rightarrow 0$ limit. The $h^2(\theta) - k(\theta)$ term
above is proportional to $V_C(\theta)$, and along with $V_N(q)$,
the remaining $q$ differentiations can produce solutions confined
arbitrarily close to a thin layer near the surface $\Sigma$.

As noted earlier, in addition to $V_C$ there is a second geometric
potential that arises from the normal part of the ${\mathbf
A}\cdot \nabla$ term in Eq. (3). Because any effective potential
on $\Sigma$ should only involve surface variables the operator
$A_N(\theta,\phi,q){\partial \over \partial q}$ must be addressed.
One prescription to deal with this operator has been given in
\cite{pershinring} and another in \cite{encinosa5} which is chosen
here. Begin by noting that the energy level spacing corresponding
to the eigenvalues for the
 $\chi_N(q)$ normal term is large compared to surface eigenvalues so it is
 safe to assume there is negligible mixing among the $\chi_N(q)$ (this assumption was
 shown to be a reasonable one in \cite{encscripta}).
 Let $G = 1+2qh(\theta)+q^2k(\theta)$ and proceed to integrate out any $q$-dependence by writing
\begin{equation}
I=\int_0^L {\chi_N(q) \over G^{1/ 2}}A_N(\theta,\phi,q)
\bigg[{\partial \over
\partial q }{\chi_N(q)\over G^{1/2}}\bigg]
G dq
\end{equation}

Given any well behaved $A_N$ it is simple to establish after an
integration by parts that the resulting effective potential
$V^{mag}$ is (with constants appended)
\begin{equation}
V^{mag}_N(\theta,\phi)={i e \hbar \over m} \bigg[
h(\theta)A_N(\theta,\phi,0)+ {1 \over 2}{\partial A_N
(\theta,\phi,q )\over
\partial q } \bigg |_{q=0} \bigg ].
\end{equation}

The ensuing Schr\"odinger equation can put into dimensionless form
by defining
$$\alpha = a/R$$
$$ F(\theta) = 1 + \rm \alpha \ cos\theta$$
$$ \gamma_N = {\pi  \hbar \over e} $$
after which $H_0$ may be written
\begin{equation}
 H_0= {\partial^2 \over \partial^2 \theta} -
  {\alpha \  {\rm sin} \ \theta \over F(\theta)}{\partial \over \partial
 \theta} + {\alpha^2 \over F^2(\theta)}{\partial^2 \over
\partial^2 \phi} + h^2(\theta)-k(\theta).
\end{equation}

The interaction potential $V(t)$ with $V^{mag}$ included becomes
$$
V(t) =  {2i \alpha^2 a_T E_0 \over \gamma_N \omega} \bigg[{\rm
sin\theta}({\rm sin \Omega  cos\phi - cos\Omega  sin\phi}){1 \over
a}{\partial \over \partial \theta}+
$$
$$
{1\over F(\theta)}{\rm ( sin\Omega sin\phi + cos\Omega\cos\phi)}
{1 \over R}{\partial \over \partial \phi}
$$
\begin{equation}
-  {1+2\alpha {\rm cos}\theta \over 2aF(\theta)} \tilde{
A_N}(\theta,\phi) -{ ka \over 4} {\rm sin} 2\theta \rm (cos\Omega
cos\phi + sin\Omega sin\phi) \bigg]
\end{equation}
where
\begin{equation}\tilde{ A_N}(\theta,\phi)= \rm -sin\Omega
cos\theta\cos\phi + \cos\Omega cos\theta \sin\phi,
\end{equation}
\begin{equation}
\Omega(\theta) = ka \ {\rm sin} \theta - \omega t
\end{equation}
and the  $|{\mathbf A}(t)|^2$ term has been neglected.

Note that the length gauge \cite{shapiro} is not employed here.
Although implementing the gauge transformation in a curved geometry
that leads to the dipole term  does not involve any inherent
difficulty, it is useful to report expressions for matrix elements
should an extension to this problem involving ionization be desired (as would
be possible for finite layer rather than ideal surface confinement), or
for cases involving thin tori with minor radii on the order of a few
angstroms subjected to visible laser light.

\section{Basis set and method}
In this work $R$ will be set to $500 \AA$, a value in accordance
with fabricated structures \cite{garsia, lorke1, lorke2,zhang},
and $\alpha = .5$, a value which serves as a compromise between
smaller $\alpha$ where the solutions tend towards simple
trigonometric functions and larger $\alpha$ which are less likely
to be physically realistic.

The eigenstates of $H_0$ are found by diagonalizing $H_0$ with a
60 state basis set expansion. The sixty states comprise five
azimuthal functions $(\nu = -2,..., 2)$  multiplied by six
positive and six negative $\theta$ parity Gram-Schmidt (GS)
functions. The GS functions are constructed to be  orthogonal over
the integration measure $ F(\theta) = 1 + \alpha \ \rm cos \theta$
\cite{encinosa6}. The energetically lowest six states are
dominated by the constant or $\rm cos \theta$ mode. To create a
non-zero net current around the minor axis of the torus, some part
of the wave function which behaves as $e^{in\theta}$ must appear;
it is not until the seventh state (as ordered by energy) that a
sine term appears which motivated the choice of the signal
frequency as $\omega_{17}=\omega_7 - \omega_1$. The resonance
frequency $\omega_{ij}$ (in natural units) between any two levels
$i$ and $j$ is numerically $3.09 \times 10^{-8} \
(\beta_j-\beta_i)\AA^{-1}$. The associated resonance wavelength
between the first and seventh states corresponds to approximately
$2 \ cm$. The electric field value chosen here is approximately
$10 \ V/m$, a value large enough to induce effects but small
enough to justify the neglect of the quadratic vector potential
term in Eq. (3). The signal was applied to the structure for ten
periods of the inverse resonance frequency, for a time of $2.04
\times 10^{10} \tau $ units, which is equivalent to $t_f = .68 ns$
with $\tau = 3 \times 10^{-11} ns$.

The matrix elements  $\big <\bar{n}\bar{\nu}| V(t)| n \nu \big
>$ can be evaluated analytically in terms of
Bessel functions \cite{matels}. The system of seven first-ordered
coupled equations is then solved for the $d_{n\nu}(t)$. Selection
rules between the $\nu$ and $\bar{\nu}$ are sufficient to render
the system of coupled equations relatively sparse allowing the
system to be solved by standard methods.

\section{Results}
In this section SCDs as computed from
\begin{equation}
\mathbf{J}(t) = {e \hbar \over m} Im \big
[\Psi^*(\theta,\phi,t)\nabla_S \Psi(\theta,\phi,t) \big ]
\end{equation}
are found. Again, the subscript on the gradient operator indicates
only surface terms are involved. From the $\mathbf{J}(t)$, magnetic
moments

\begin{equation}
\mathbf{M}(t) = {1 \over 2} \int_0^{2\pi}\int_0^{2\pi} \mathbf{r}
\times \mathbf{J} dA
\end{equation}
with $dA = aW(\theta)d\theta d\phi$  are presented.  It is
possible to write $\mathbf{M}(t)$ in terms of unit vectors
${\mathbf e}_{\theta},{\mathbf e}_{\phi},{\mathbf e}_{n}$, but
because  the interest lies in comparing solenoidal to dipole
modes, the components along ${\mathbf e}_{\rho},{\mathbf
e}_{\phi},{\mathbf e}_{z}$  are shown instead. Explicit forms for
the magnetic moments $M_\rho (t), M_\phi (t), M_z (t)$ are given
in the Appendix.

Before presenting results for the currents and moments, it is worth
showing representative plots of the time dependent coefficients
$d_i(t)$. First let $L$ serve as a collective index for the values
$i=1....6$. In Fig. 1, $|d_1(t)|^2$ and $|d_7(t)|^2$ for the LPW are
shown, and Fig. 2 gives the same for the CPW. In both situations,
$|d_7(t)|^2$ is small as compared to the remaining $|d_L(t)|^2$, but
oscillate at a much slower frequency than the $|d_L(t)|^2$.  As was
noted previously, it is $d_7(t)$ that must multiply the sine terms
to combine with the positive $\theta$ parity parts of the total wave
function to yield currents around the minor radius, so it is the
time scale of $d_7(t)$ that will set the time scale and magnitudes
of $\mathbf{J}(t)$ and $\mathbf{M}(t)$.

Results for SCDs as functions of $\theta$ evaluated at $t_f$ are
shown in Figs. 3 and 4 for the LPW. Fig. 3 plots
$J^a_\theta(t_f,\theta)$, the current calculated from employing only
the $d_L(t)$, and Fig. 4 plots $J^b_\theta(t_f,\theta)$, the current
resulting from inclusion of $d_7(t)$ summed over the $d_L(t)$, both
at $\phi = 0 $. The results illustrate that the net current
resulting from $J^a_\theta(t_f,\theta)$ is zero, but is non-zero for
$J^b_\theta(t_f,\theta)$. Figs. 5 and 6 plot the same quantities for
the CPW. Results for the azimuthal current $J_\phi(t_f,\phi)$ at
$\theta = 0$ and $\theta = \pi$  are shown in Figs. 7  for the LPW
and in Fig. 8 for the CPW. The contribution from terms proportional
to $d_7(t)$ is always much smaller than the contribution arising
solely from the $d_L(t)$ terms, so they have not been shown
separately. No net azimuthal current results from the LPW, and only
a small current is present in for the CPW case.

In Fig. 9 the $M_i(t)$ for the LPW are shown for a duration of
$t_f$ at
 $\phi = \pi / 2, \theta = 0$. The surprising result here is that
$M_\phi(t)$ is generally an order of magnitude larger than the
dipole moment $M_z(t)$. Although not shown here, it is observed
that as the torus is traversed in the $\phi$ direction, the
direction of the largest component of the magnetic moment is
consistent with a  magnetic moment perpendicular to the direction
of polarization. The corresponding results for the CPW presented
in Fig. 10 indicate that the dipole mode here is four orders of
magnitude larger than any solenoidal modes, a result that holds
true at every angle.

The above results were applicable to an ideal situation wherein
the structure is situated on a surface transparent to microwaves.
It is interesting to consider what is perhaps the more realistic
case of the incident wave interfering with its reflected part.

In order to see whether the interference of two  polarized waves
propagating along the z-axis
 can enhance transitions to  solenoidal
modes,
 an incoming electromagnetic wave being
reflected at a completely reflecting flat surface ($ \lq$mirror')
parallel to the horizontal symmetry plane of the torus was
considered. The mirror is positioned at variable distances $z_0$
from the central plane in order to study how superposition of the
incoming and reflected linear polarized wave fronts might affect
the relative amplitude of the solenoidal modes. The general
arguments will be made for the CPW; the LPW results are a
special case of those for the CPW.

At the mirror, the electric field must satisfy the boundary
condition of no total transverse modes
\begin{equation}
{\mathbf E}_{in}(z_0,t)+{\mathbf E}_{out}(z_0,t)=0,
\end{equation}
The location and effect of the reflecting mirror on the incoming
wave front in this context may be simply modeled in form of an
additional phase $\phi_0=-2 k z_0$ in the reflected field
amplitude. The signs of $\omega$ and $z_0$ can be chosen according
from the incident direction of the incoming wave.  From Eq. (25), the reflected wave amplitude must be:
\begin{equation}
{\mathbf E}_{out}(z,t)= {E_0} \big[-{\rm cos}(k z +\omega t ')
{\mathbf i}+{\rm sin}(k z +\omega t ') {\mathbf j}\big]
\end{equation}
with $ t' = t-t_0$; and $ t_0 = {- {2 k z_0}/\omega}$.
Thus, the reflected wave amplitude  can be obtained from the
incoming wave amplitude by letting
\begin{equation}
\label{A_circ_intf_sub} \omega \rightarrow -\omega, \,\, \, t
\rightarrow t', \,\, \, {\mathbf j} \rightarrow -{\mathbf j}.
\end{equation}
in Eq. (26).

Several different positions for the position of the mirror plane
($z_0=a$, $\lambda\over 4$, $\lambda\over 2$ and $\lambda\over
8$), were investigated. The general outcome is simple oscillatory
behavior with different but small amplitudes while
$\big|d_1\big|^2$ stays close to 1. For $z_0={\lambda\over 4}$,
all coefficients $d_2, \dots, d_7$ are numerically nearly equal to
zero, while $d_1$ stays constant and identical to 1. This is due
to the fact that the electric field amplitude ${\mathbf E}$
reaches its maximum value, thus the field gradient is essentially
zero over the whole torus. In all of these cases, the system stays
essentially in the ground state and no transitions are observed.

The optimal mirror position that maximized the value
$\big|d_7(t)\big|^2$ was determined to be at
\begin{equation}
z_0 \approx 0.0027 \lambda,
\end{equation}
where it was found that $\big|d_7\big|^2$ reaches its maximum
amplitude of roughly $7*10^{-7}$.  This is an increase of up to 3
to 4 orders of magnitude for positions around $\lambda\over 4$
 discussed above.

 The analogous results for the interference cases to those
 presented above are given below. Figs. 11 and 12 show
$|d_1(t)|^2$ and $|d_7(t)|^2$ plots for the LPW and CPW. It is
worth noting that while the $|d_7(t)|^2$ time scale does not vary
greatly due to interference as compared to the results in Figs. 1
and 2, the $|d_1(t)|^2$ time scale is very different in both
cases.

The SCD plots for the LPW do not appear qualitatively different
than the results already presented in Figs. 3, 4 and 7, so in the
interest of conciseness those results are not presented here.
However, an interesting result is shown in Fig. 13, wherein a
clear signature of a circulatory $J^b_\theta(t_f,\theta)$ SCD at
$\phi = 0$ is shown for the CPW due to the presence of the
$d_7(t)$ term. Fig. 14 shows there is also a circulatory azimuthal
SCD at $\theta = \pi$.

The LPW magnetic moment results are given in Fig. 15. The
frequency of oscillation has again been reduced, and the overall
magnitude of $M_z(t)$ is smaller than that of $M_\phi(t)$. Fig. 16
shows that for the CPW, $M_z(t)
> M_\phi(t)$; this is observed to be independent of the azimuthal
point at which the $M_i(t)$ are calculated.

\section{Conclusions}

This work presents a general framework for calculating surface
current densities and magnetic moments on a toroidal surface in
the presence of an electromagnetic wave. A proof of principle
calculation demonstrating that polarized microwaves can cause
circulating surface currents around the toroidal minor radius in
addition to azimuthal currents was given. Rather than employ the
dipole approximation, matrix elements of the electromagnetic term
were evaluated in closed form. While there can be little doubt that
the dipole approximation would be perfectly valid for the system
considered, there are cases involving ionization or surface states
with energies characterized by much smaller minor radii that
warrant exact expressions.

The magnetic moments calculated in this work show that it is in
principle possible to manipulate surface currents in a manner that
causes the moments to $ \lq\lq$point" predominantly in particular
directions at certain times.  Our preliminary results indicate
that interference due to reflection could potentially play an
important role in the development of this topic.

The realization of the model problem considered here may be
physically realizable with current experimental methods by placing
a thin layer of a good conductor over an InAs toroidal structure.
The extension of this work to metallic carbon nanotube tori is
likely possible but would require some effort to account for the
lattice. Preliminary work indicates that modelling the carbon
sites by weak delta function potentials requires a larger basis
set than employed here.

\newpage

\begin{center} {\bf Acknowledgments} \end{center} The authors would like to thank B. Etemadi for useful discussions.  This work has been funded in parts by NIH Grant HD 41829 (M. Jack).

\newpage

\renewcommand{\theequation}{A.\arabic{equation}}
  % redefine the command that creates the equation no.
\setcounter{equation}{0}  % reset counter 

\section*{Appendix}  % use *-form to suppress numbering

This appendix presents closed-form expressions for the integrated magnetic moment and its components.  The magnetic moment ${\bf M}(t)$ has been deferred as:
\begin{equation}
\label{mag_mu}
%{\bf M}(t) =\int\limits_{S} d{\sigma} \,\, {\mathbf r} \times {\mathbf J({\mathbf r}, t)}
\mathbf{M}(t) = {1 \over 2} \int_0^{2\pi}\int_0^{2\pi} \mathbf{r}
\times \mathbf{J}({\mathbf r}, t) dA
\end{equation}
with the radius vector ${\mathbf r}$ on the $T^2$ surface generated by
\begin{eqnarray}
\label{radius_r}
{\mathbf r}  \equiv {(R+a\cos\theta)} {{\mathbf e}_\rho}  + {a\sin\theta}{{\mathbf e}_z }
\end{eqnarray}
and the surface gradient being
\begin{eqnarray}
\label{dr_nabla_ds}
\nabla_S = {{\mathbf e}_\theta} \, {1 \over a}\,  {{\partial}\over{\partial\theta}} 
+{{\mathbf e}_\phi}\,  {1 \over {R+a\cos\theta}}\,  {{\partial}\over{\partial\phi}}.
\end{eqnarray}
The time dependent current density ${\mathbf J} ({\mathbf r}, t)$ in Eq. (A.1) is defined in Eq. (23) with the wave function $\Psi ({\mathbf r}, t)$ given in Eq. (2) as solution of the time dependent Schr\"odinger equation.  Here express the $\theta$ dependent part of the eigenstates, $\chi_n(\theta)$, directly in terms of sums of cosines with coefficients  $C_n^k$ for positive $\theta$ parity solutions, or as sums of sines with coefficients  $D_n^k$ for negative $\theta$ parity solutions respectively:
\begin{equation}
\label{eigenstates}
\chi_n(\theta) 
=
\left\{
\begin{array}{lll}
{ 1 \over \sqrt{\pi} } \sum\limits_{k=0}^{5} C_n^k \cos(k\theta)
&
\textrm{for}
& n=1\ldots 6;
\\
{ 1 \over \sqrt{\pi} } \sum\limits_{k=1}^{5} D_n^k \sin(k\theta)
&
\textrm{for}
& n=7.
\end{array}
\right.
\end{equation}
Now introduce the following constants $M_0$, $C$ and $A$:
\begin{eqnarray}
\label{constant_C}
M_0 = {{e \hbar }\over {4 m_e}} R^2;\qquad C = {{e \hbar }\over {2 m_e}};\qquad A = {\pi R^2}.
\end{eqnarray}
Applying Eq. (A.2) through Eq. (A.5) the time dependent magnetic moment ${\bf M}(t)$ in Eq. (A.1) can be written as integral via the surface variables $\theta$ and $\phi$ with $F(\theta)\equiv 1+\alpha\cos\theta$:
\begin{eqnarray}
\label{mag_mu_new}
{\bf M}(t) &=& -i  {\pi M_0} \int\limits_{0}^{2\pi}d{\phi} \int\limits_{-\pi}^{\pi}d{\theta} 
\Big[
\alpha F(\theta) {J_\phi}\,  {{\mathbf e}_z}  
-
F(\theta)(\cos\theta+\alpha) {J_\theta} \, {{\mathbf e}_\phi}  
-
\alpha^2\sin\theta {J_\phi}\, {{\mathbf e}_\rho}
\Big]
\end{eqnarray}
with following expressions for the current densities:
\begin{eqnarray}
\label{currents_jth_jph}
{J_\phi}({\mathbf r}, t)  &= &
\Psi^{*} ({\mathbf r}, t)  { {\partial \Psi } \over  {\partial \phi} }({\mathbf r}, t)
-  
\Psi ({\mathbf r}, t)  { {\partial \Psi^{*} } \over  {\partial \phi} }({\mathbf r}, t) 
; \\
{J_\theta}({\mathbf r}, t) &= &
\Psi^{*} ({\mathbf r}, t)  { {\partial \Psi } \over  {\partial \theta} }({\mathbf r}, t)
-  
\Psi ({\mathbf r}, t)  { {\partial \Psi^{*} } \over  {\partial \theta} }({\mathbf r}, t)
.
\end{eqnarray}
The currents in Eq. (A.7) and Eq. (A.8) can be re-expressed as
\begin{eqnarray}
\label{currents_jth_jph_1}
{J_\phi} ({\mathbf r}, t)& = &
\sum\limits_{i,j} d_i^{*}(t) d_j(t) e^{ i \omega_{ij}  t} {J_\phi^{ij}}({\mathbf r})
,\\
{J_\theta} ({\mathbf r}, t) &=& 
\sum\limits_{i,j} d_i^{*}(t) d_j(t) e^{i \omega_{ij}  t} {J_\theta^{ij}}({\mathbf r})
,
\end{eqnarray}
with the time independent current expressions ${J_\phi^{ij}} (\theta,\phi)$ and ${J_\theta^{ij}} (\theta,\phi)$ presented in the following way ($\omega_{ij} \equiv  {E_{ij} \over \hbar }\equiv  {{ E_i - E_j } \over \hbar }$; $\nu_{ij} \equiv  \nu_i - \nu_j$):
\begin{eqnarray}
\label{currents_jth_jph_2}
{J_\phi^{ij}} (\theta,\phi) 
&=& 
2 i \left(\nu_i +\nu_j\right) e^{- i \nu_{ij}\phi } 
\chi_i(\theta) \chi_j(\theta)  
;\\
{J_\theta^{ij}} (\theta,\phi) 
&=& 
2 e^{- i \nu_{ij}\phi }
\Big(
\chi_i^{*} { {\partial \chi_j} \over  {\partial \theta} }
-  
\chi_j  { {\partial \chi_i^{*} } \over  {\partial \theta} } 
\Big)
.
\end{eqnarray}

For the final results for the magnetic moments we prepare the integrations over $\theta$ and $\phi$.  The $\phi$-integration yields following three integrals:
\begin{eqnarray}
\label{integrals_phi}
\int\limits_{0}^{2\pi}d{\phi} \, e^{ i (\omega_{ij}  t - \nu_{ij}\phi) } {\mathbf e}_z 
&=& 
2\pi\delta(\nu_i-\nu_j) e^{ i \omega_{ij}  t} {\mathbf e}_z
,\\
\int\limits_{0}^{2\pi}d{\phi} \, e^{ i (\omega_{ij}  t - \nu_{ij}\phi) } {\mathbf e}_\phi  
&=& 
\pi
\bigg[ 
\Big( \delta_p  + \delta_n \Big) {\mathbf e}_y 
+ i \Big( \delta_p  - \delta_n \Big) {\mathbf e}_x  
\bigg]
e^{ i \omega_{ij}  t} 
 ,\\
\int\limits_{0}^{2\pi}d{\phi} \, e^{ i (\omega_{ij}  t - \nu_{ij}\phi) } {\mathbf e}_\rho
&=& 
\pi
\bigg[ 
\Big( \delta_p  + \delta_n \Big) {\mathbf e}_x 
- i \Big( \delta_p  - \delta_n \Big) {\mathbf e}_y  
\bigg]
e^{ i \omega_{ij}  t}
, 
\end{eqnarray}
which includes the Kroneckerdelta expressions
\begin{eqnarray}
\label{kroneckerdeltas_phi}
\delta_p \equiv \delta(\nu_j-\nu_i+1);& \delta_n \equiv \delta(\nu_j-\nu_i-1);
\\
\label{kroneckerdeltas_theta}
\delta_{-} \equiv \delta(k-l); &\delta_{+} \equiv \delta(k+l);
\\
\label{kroneckerdeltas_theta_1}
\delta_{-}^{\pm 1}\equiv \delta(k-l \pm 1); &\delta_{+}^{\pm 1}\equiv \delta(k + l \pm 1);
\\
\label{kroneckerdeltas_theta_2}
\delta_{-}^{\pm 2}\equiv \delta(k-l \pm 2); &\delta_{+}^{\pm 2}\equiv \delta(k + l \pm 2).
\end{eqnarray}
The final results for the $\theta$-integration are based on following integrals of products of sines and cosines:
\begin{eqnarray}
\label{integrals_theta_1}
\int\limits_{-\pi}^{\pi}d{\theta} \, \alpha F(\theta) \cos(k\theta) \cos(l\theta) = 
{ {\alpha}\over 2} 
\Bigg[
\delta_{-} + \delta_{+} 
+ { {\alpha}\over 2} 
\Big(
\delta_{-}^{-1} + \delta_{-}^{+1} +\delta_{+}^{-1} + \delta_{+}^{+1} 
\Big)
\Bigg]
;\\
\int\limits_{-\pi}^{\pi}d{\theta} \, \alpha F(\theta) \sin(k\theta) \sin(l\theta) = 
{ {\alpha}\over 2} 
\Bigg[
\delta_{-} - \delta_{+} 
+ { {\alpha}\over 2} 
\Big(
\delta_{-}^{-1} + \delta_{-}^{+1} -\delta_{+}^{-1} - \delta_{+}^{+1} 
\Big)
\Bigg]
;
\end{eqnarray}
\begin{eqnarray}
\label{integrals_theta_2}
&&\int\limits_{-\pi}^{\pi}d{\theta} \,  F(\theta) (\cos\theta+\alpha) \cos(k\theta) \cos(l\theta) =
\nonumber\\ 
& =&
{ 1 \over 2} 
\Bigg[
{ 3\over 2} \alpha
\Big(
\delta_{-} + \delta_{+} 
\Big)
+ { 1 \over 2} 
\Big(
1+\alpha^2
\Big)
\Big(
\delta_{-}^{-1} + \delta_{-}^{+1} +\delta_{+}^{-1} + \delta_{+}^{+1} 
\Big)
+ { {\alpha}\over 4} 
\Big(
\delta_{-}^{-2} + \delta_{-}^{+2} +\delta_{+}^{-2} + \delta_{+}^{+2} 
\Big)
\Bigg]
;
\nonumber\\
\\
&&\int\limits_{-\pi}^{\pi}d{\theta} \,  F(\theta) (\cos\theta+\alpha) \sin(k\theta) \sin(l\theta) =
\nonumber\\
& = &
{ 1 \over 2} 
\Bigg[
{ 3\over 2} \alpha
\Big(
\delta_{-} - \delta_{+} 
\Big)
+ { 1 \over 2} 
\Big(
1+\alpha^2
\Big)
\Big(
\delta_{-}^{-1} + \delta_{-}^{+1} -\delta_{+}^{-1} - \delta_{+}^{+1} 
\Big)
 + { {\alpha}\over 4} 
\Big(
\delta_{-}^{-2} + \delta_{-}^{+2} -\delta_{+}^{-2} -\delta_{+}^{+2} 
\Big)
\Bigg]
;
\nonumber\\
\end{eqnarray}
\begin{eqnarray}
\label{integrals_theta_3}
\int\limits_{-\pi}^{\pi}d{\theta} \, \alpha^2 \sin(\theta) \cos(k\theta) \sin(l\theta) = 
{ {\alpha^2}\over 4} 
\Big(
\delta_{+}^{-1} - \delta_{+}^{+1} -\delta_{-}^{-1} + \delta_{-}^{+1} 
\Big)
.
\end{eqnarray}
These results in Eq. (A.20) to (A.24) combine to the final $\theta$-integrals of products of eigenstates $\chi_i$, $\chi_j$:
\begin{eqnarray}
\label{mag_mu_1_phi}
\int\limits_{-\pi}^{\pi} d{\theta}\, \chi_i(\theta)\, \chi_j(\theta)\, F(\theta) 
&=&
{{1} \over {2 \pi}}\left\{
\begin{array}{lll}
\sum\limits_{k,l=0}^{5} C_i^k C_j^l 
\bigg[
\delta_{-} + \delta_{+} 
+ { {\alpha}\over 2} 
\Big(
\delta_{-}^{-1} + \delta_{-}^{+1} +\delta_{+}^{-1} + \delta_{+}^{+1} 
\Big)
\bigg]
&
\textrm{for}
& i,j = 1\ldots 6;
\\
&
\\
\sum\limits_{k,l=1}^{5} D_i^k D_j^l 
\bigg[
\delta_{-} - \delta_{+} 
+ { {\alpha}\over 2} 
\Big(
\delta_{-}^{-1} + \delta_{-}^{+1} -\delta_{+}^{-1} - \delta_{+}^{+1} 
\Big)
\bigg]
&
\textrm{for}
& i=j=7;
\end{array}
\right.
\nonumber\\
\end{eqnarray}
\begin{eqnarray}
\label{mag_mu_2_phi}
&&\int\limits_{-\pi}^{\pi} d{\theta}
\Big(
\chi_i { {\partial \chi_j} \over  {\partial \theta} }
-  
\chi_j  { {\partial \chi_i} \over  {\partial \theta} } 
\Big)
F(\theta)\, (\cos\theta + \alpha)
=
\nonumber\\ 
&=&
{{1} \over {2 \pi} }\sum\limits_{k=0}^{5} 
\sum\limits_{l=1}^{5} 
\Bigg\{
k\cdot
\bigg[
{ 3\over 2} \alpha
\Big(
\delta_{-} + \delta_{+} 
\Big)
+ { 1 \over 2} 
\Big(
1+\alpha^2
\Big)
\Big(
\delta_{-}^{-1} + \delta_{-}^{+1} +\delta_{+}^{-1} + \delta_{+}^{+1} 
\Big)
\nonumber\\
&& 
\qquad\qquad\qquad\quad
+ { {\alpha}\over 4} 
\Big(
\delta_{-}^{-2} + \delta_{-}^{+2} +\delta_{+}^{-2} + \delta_{+}^{+2} 
\Big)
\bigg]
\nonumber\\
&&
\qquad\qquad\quad
+
l\cdot 
\bigg[
{ 3\over 2} \alpha
\Big(
\delta_{-} - \delta_{+} 
\Big)
+ { 1 \over 2} 
\Big(
1+\alpha^2
\Big)
\Big(
\delta_{-}^{-1} + \delta_{-}^{+1} -\delta_{+}^{-1} - \delta_{+}^{+1} 
\Big)
\nonumber\\
&&
\qquad\qquad\qquad\quad
+ { {\alpha}\over 4} 
\Big(
\delta_{-}^{-2} + \delta_{-}^{+2} -\delta_{+}^{-2} - \delta_{+}^{+2} 
\Big)
\bigg]
\Bigg\}
\nonumber\\
&&
\qquad\qquad\quad
\cdot \left\{
\begin{array}{lllll}
C_i^k D_j^l 
&
\textrm{for}
& i= 1\ldots 6
&
\textrm{and}
&
j=7;
\\
\left(-C_j^k D_i^l\right) 
&
\textrm{for}
& i=7
&
\textrm{and}
&
j=1\ldots 6;
\end{array}
\right.
\nonumber\\
\end{eqnarray}
\begin{eqnarray}
\label{mag_mu_3_phi}
&&\int\limits_{-\pi}^{\pi} d{\theta}
\chi_i(\theta) \, \chi_j(\theta) \sin\theta
=
\nonumber\\ 
&=&
{1 \over {4 \pi}}\sum\limits_{k=0}^{5} 
\sum\limits_{l=1}^{5} 
\Big(
\delta_{+}^{-1} - \delta_{+}^{+1} -\delta_{-}^{-1} + \delta_{-}^{+1} 
\Big)
\cdot 
\left\{
\begin{array}{lllll}
C_i^k D_j^l 
&
\textrm{for}
& i= 1\ldots 6
&
\textrm{and}
&
j=7;
\\
C_j^k D_i^l
&
\textrm{for}
& i=7
&
\textrm{and}
&
j=1\ldots 6.
\end{array}
\right.
\nonumber\\
\end{eqnarray}
In the final expression for the magnetic moment ${\bf M}$, re-express ${\mathbf e}_x$ and ${\mathbf e}_y$ 
in terms of ${\mathbf e}_\phi$ and ${\mathbf e_\rho}$ while introducing the fixed angular variable $\phi$:
\begin{eqnarray}
\label{unit_vectors}
{\mathbf e}_x
&=& 
- \sin\phi \, {\mathbf e}_\phi + \cos\phi \, {\mathbf e}_\rho;
\nonumber\\
{\mathbf e}_y
&=& 
\cos\phi \, {\mathbf e}_\phi + \sin\phi \, {\mathbf e}_\rho
.
\end{eqnarray}
The $\theta$-integration finally yields three contributions ${\bf M}_1(t)$, ${\bf M}_2(t)$ and ${\bf M}_3(t)$ for the total magnetic moment.  The dipole mode ${\bf M}_1$ ($\equiv {\bf M}_z$) parallel to the central symmetry axis can be determined using the definitions $d_i^R(t) \equiv \textrm{Re}\left(d_i(t)\right) $ and $d_i^I(t) \equiv \textrm{Im}\left(d_i(t)\right)$ as:
\begin{eqnarray}
\label{mag_mu_1}
{\bf M}_1 (t) \equiv {\bf M}_z (t) &=& -i  {\alpha \pi M_0}  \int\limits_{0}^{2\pi} d{\phi} \int\limits_{-\pi}^{\pi} d{\theta}
F(\theta) {J_\phi} (\theta,\phi,t)\,  {{\mathbf e}_z} 
= 
\nonumber\\
&=&
 {\mathbf e}_z \,  
(4 \alpha \pi M_0)
\sum\limits_{i,j=1}^{7}  
\left(
\nu_i+\nu_j
\right)
\delta\left(
\nu_i-\nu_j
\right)
\nonumber\\
&&
\cdot 
\bigg[
\Big(
d_i^R(t) d_j^R(t) + d_i^I(t) d_j^I(t) 
\Big)
\cos(\omega_{ij}t)
-
\Big(
d_i^R(t) d_j^I(t) - d_i^I(t) d_j^R(t) 
\Big)
\sin(\omega_{ij}t)
\bigg]
\nonumber\\
&& 
\int\limits_{-\pi}^{\pi} d{\theta} \, \chi_i(\theta)\, \chi_j(\theta)\, F(\theta)
;
\qquad\qquad\qquad
4 \alpha \pi M_0 = {{e \hbar }\over {2 m_e}} (2\pi a R).
\end{eqnarray}

The two solenoidal modes are expressed by ${\bf M}_2(t)$ and ${\bf M}_3(t)$.  ${\bf M}_2$ is proportional to the integrated current term $J_\theta$ in Eq. (A.8):

\begin{eqnarray}
\label{mag_mu_2}
{\bf M}_2 (t) &=& i {\pi M_0} \int\limits_{0}^{2\pi} d{\phi} \int\limits_{-\pi}^{\pi} d{\theta}
F(\theta)(\cos\theta+\alpha) {J_\theta} (\theta,\phi,t) \, {{\mathbf e}_\phi}  
=
\nonumber\\
&=& 
- 2 \pi M_0
\sum\limits_{i,j=1}^{7}  
\Bigg\{
\bigg[
\Big(
d_i^R(t) d_j^I(t) - d_i^I(t) d_j^R(t) 
\Big)
\Big(
\delta_p\cos(\omega_{ij}t-\phi) +\delta_n\cos(\omega_{ij}t+\phi) 
\Big)
\nonumber\\
&&
\qquad\qquad
+
\Big(
d_i^R(t) d_j^R(t) + d_i^I(t) d_j^I(t) 
\Big)
\Big(
\delta_p\sin(\omega_{ij}t-\phi) +\delta_n\sin(\omega_{ij}t+\phi) 
\Big)
\bigg]
{\mathbf e}_\phi
\nonumber\\
&&
\qquad\qquad
+ 
\bigg[
\Big(
d_i^R(t) d_j^I(t) - d_i^I(t) d_j^R(t) 
\Big)
\Big(
\delta_p\sin(\omega_{ij}t-\phi) -\delta_n\sin(\omega_{ij}t+\phi) 
\Big)
\nonumber\\
&&
\qquad\qquad
+
\Big(
d_i^R(t) d_j^R(t) + d_i^I(t) d_j^I(t) 
\Big)
\Big(
\delta_p\cos(\omega_{ij}t-\phi) -\delta_n\cos(\omega_{ij}t+\phi) 
\Big)
\bigg]
{\mathbf e}_\rho
\Bigg\}
\nonumber\\
&& 
\qquad\qquad
\cdot \int\limits_{-\pi}^{\pi} d{\theta} 
\Big(
\chi_i { {\partial \chi_j } \over  {\partial \theta} }
-  
\chi_j  { {\partial \chi_i } \over  {\partial \theta} } 
\Big)
F(\theta) \, (\cos\theta + \alpha)
;
\quad
2 \pi M_0 = {{e \hbar }\over {2 m_e}} (\pi R^2).
\end{eqnarray}
The second solenoidal mode ${\bf M}_3$ is proportional to the integrated current $J_\phi$ in Eq. (A.7):
\begin{eqnarray}
\label{mag_mu_3}
&& {\bf M}_3 (t)= i  {\alpha^2 \pi M_0} \int\limits_{0}^{2\pi} d{\phi} \int\limits_{-\pi}^{\pi} d{\theta}
\sin\theta {J_\phi} (\theta,\phi,t)\, {{\mathbf e}_\rho}
=
\nonumber\\
&=& 
-4 {\alpha^2} \pi M_0
\sum\limits_{i,j=1}^{7} 
{\left(\nu_i+\nu_j\right) \over 2}
\Bigg\{
\bigg[
\Big(
d_i^R(t) d_j^I(t) - d_i^I(t) d_j^R(t) 
\Big)
\Big(
\delta_p\cos(\omega_{ij}t-\phi) -\delta_n\cos(\omega_{ij}t+\phi) 
\Big)
\nonumber\\
&&
\qquad\qquad\qquad\qquad
+
\Big(
d_i^R(t) d_j^R(t) + d_i^I(t) d_j^I(t) 
\Big)
\Big(
\delta_p\sin(\omega_{ij}t-\phi) -\delta_n\sin(\omega_{ij}t+\phi) 
\Big)
\bigg]
{\mathbf e}_\phi
\nonumber\\
&&
\qquad\qquad\qquad\qquad
+ 
\bigg[
-\Big(
d_i^R(t) d_j^I(t) - d_i^I(t) d_j^R(t) 
\Big)
\Big(
\delta_p\sin(\omega_{ij}t-\phi) + \delta_n\sin(\omega_{ij}t+\phi) 
\Big)
\nonumber\\
&&
\qquad\qquad\qquad\qquad
+
\Big(
d_i^R(t) d_j^R(t) + d_i^I(t) d_j^I(t) 
\Big)
\Big(
\delta_p\cos(\omega_{ij}t-\phi) + \delta_n\cos(\omega_{ij}t+\phi) 
\Big)
\bigg]
{\mathbf e}_\rho
\Bigg\}
\nonumber\\
&& 
\qquad\qquad\qquad\qquad
\cdot \int\limits_{-\pi}^{\pi} d{\theta} 
\, \chi_i(\theta) \, \chi_j(\theta) \sin\theta
;
\qquad\qquad\qquad
2 {\alpha^2} \pi M_0 =  {{e \hbar }\over {2 m_e}} \left(\pi a^2\right).
\end{eqnarray}
The two components mentioned in the text, $M_\rho(t)$ and $M_\phi(t)$, can then be very easily obtained from Eq. (A.30) and (A.31) by adding ${\bf M}_2$ and ${\bf M}_3$ and projecting into the directions ${\mathbf e}_\rho$ or ${\mathbf e}_\phi$.

\newpage

%\begin{center} {\bf Figures} \end{center}

%\begin{figure}
%\centering
%\includegraphics{Figure_1.eps}
%\caption{insert caption here}
%\vskip 16pt %\centerline{Fig. 1, Encinosa and Jack, P and N}
%\end{figure}

\bibliography{refcomp3}

\newpage

\begin{center} {\bf Figures} \end{center}

\begin{figure}[h]
\centering
\includegraphics{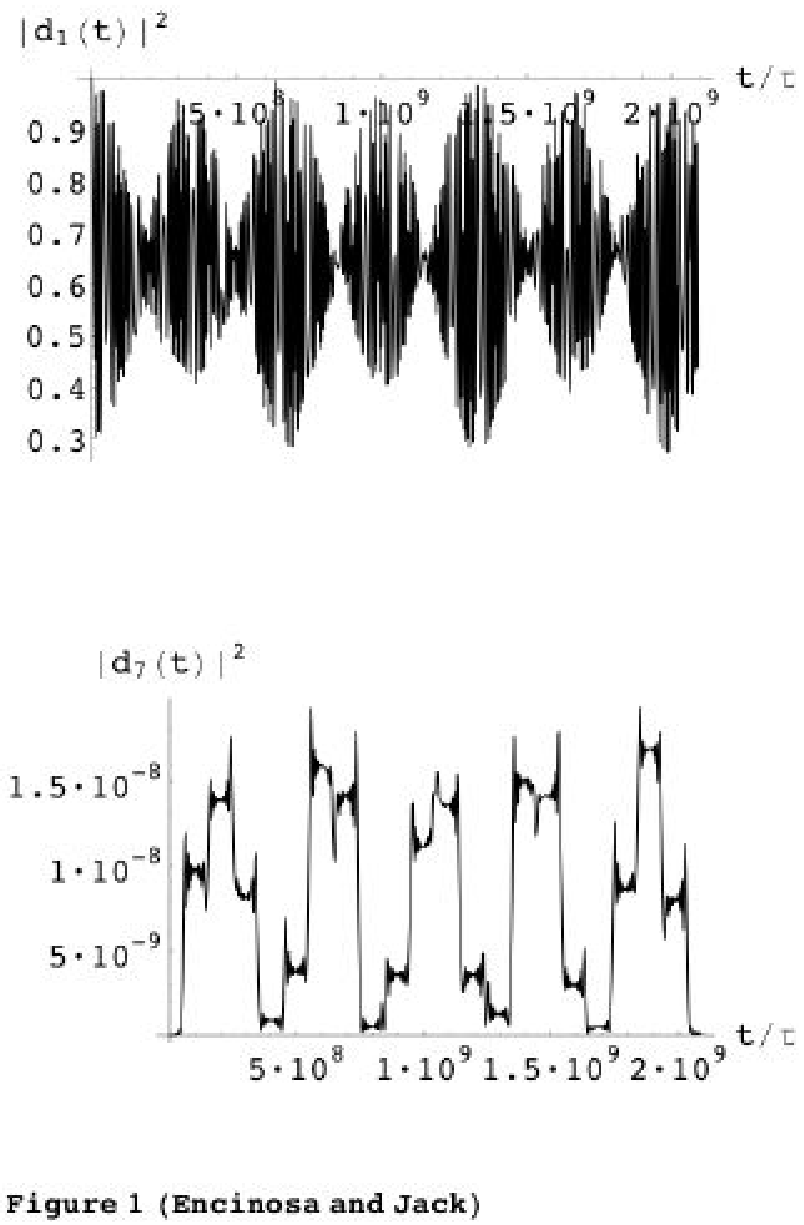}
\caption{Ground state $|d_1(t)|^2$ and
seventh  state $|d_7(t)|^2$ probability amplitudes  for the system
subjected to a LPW along the toroidal symmetry axis (time scale here: $\tau = 3 \times 10^{-11}\,ns$).}
%\centerline{Fig. 1}
\end{figure}
\begin{figure}[h]
\centering
\includegraphics{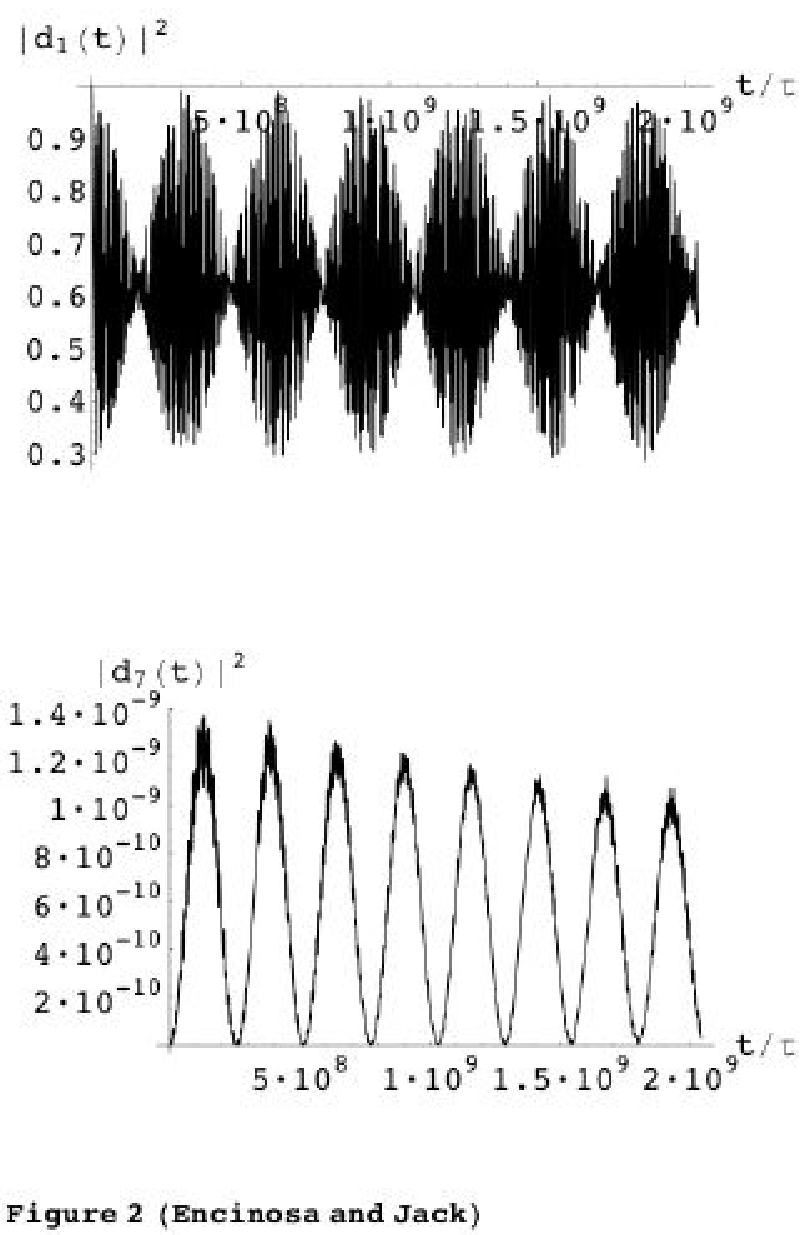}
\caption{Ground state $|d_1(t)|^2$ and
seventh  state $|d_7(t)|^2$ probability amplitudes  for the system
subjected to a CPW along the toroidal symmetry axis.}
%\centerline{Fig. 2}
\end{figure}

%\newpage

\begin{figure}
\centering
\includegraphics{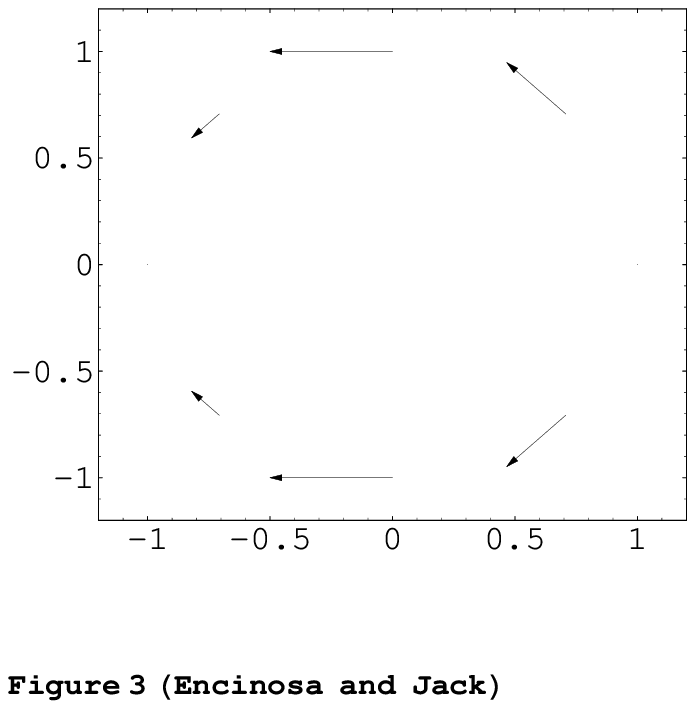}
\caption{$J^a_\theta(t_f,\theta)$ in units
of $J_0=e \hbar / m$ (longest arrow = $.00015$) plotted on the torus (in units
of $R = 1$) at $\phi = 0$ in $\theta = \pi/4$ intervals for the
LPW case. The net current integrated over the loop is zero ($t_f = 0.68 ns$).}
%\centerline{Fig. 3}
\end{figure}
\begin{figure}
\centering
\includegraphics{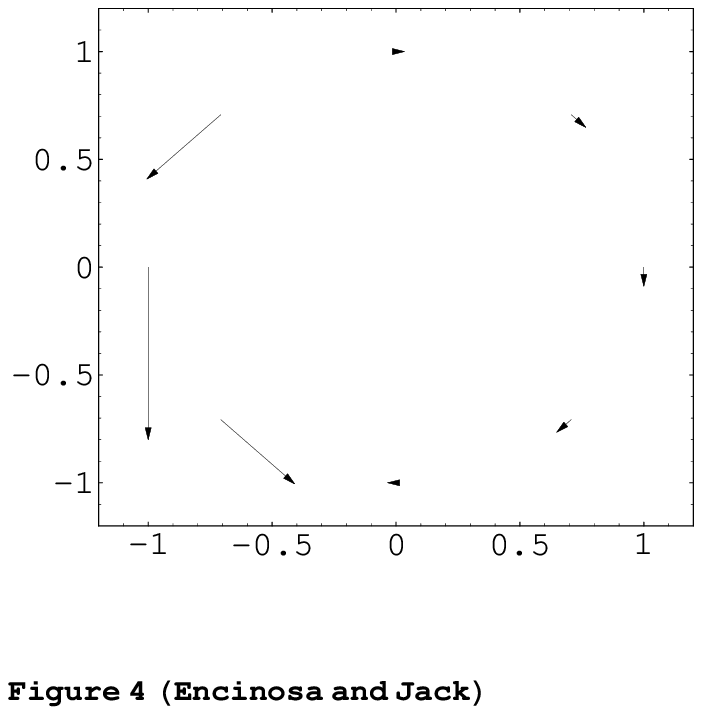}
\caption{$J^b_\theta(t_f,\theta)$ plotted
on the torus (in units of $R = 1$) at $\phi = 0$ in $\theta =
\pi/4$ intervals for the LPW case. The longest arrow is $3 \times
10^{-6}J_0$. The net current integrated over the loop is non-zero
and arises from the presence of the negative $\theta$ parity
$d_7(t)$ amplitude.}
%\centerline{Fig. 4}
\end{figure}

%\newpage

\begin{figure}
\centering
\includegraphics{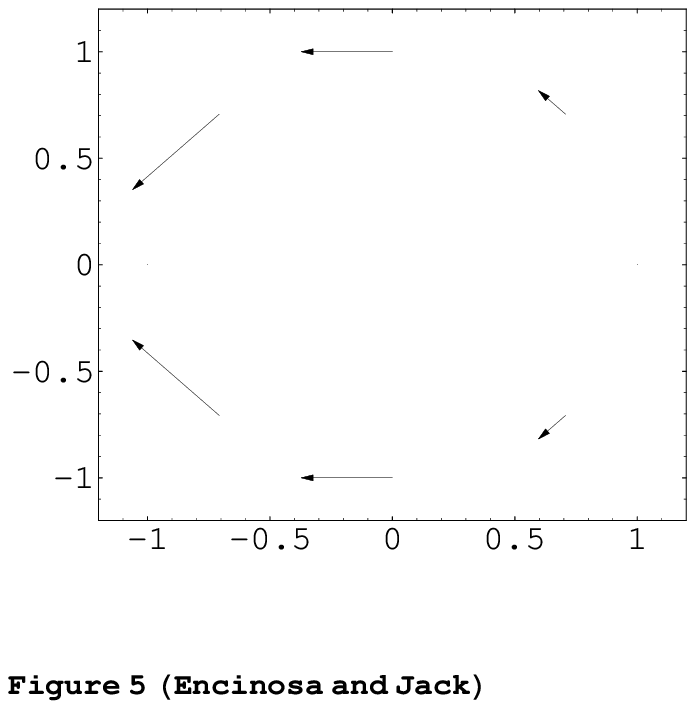}
\caption{$J^a_\theta(t_f,\theta)$ plotted
on the torus (in units of $R = 1$) at $\phi = 0$ in $\theta =
\pi/4$ intervals for the CPW case. The longest arrow is $.034
\times 10^{-6}J_0$. The net current integrated over the loop is
zero.}
%\centerline{Fig. 5}
\end{figure}
\begin{figure}
\centering
\includegraphics{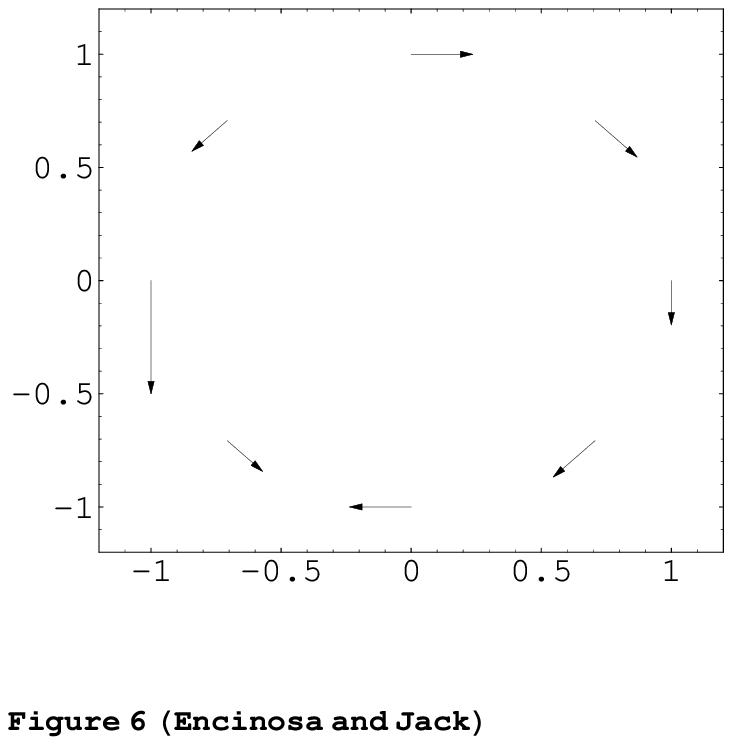}
\caption{$J^b_\theta(t_f,\theta)$ plotted
on the torus (in units of $R=1$) at $\phi = 0$ in $\theta = \pi/4$
intervals for the CPW case. The longest arrow is $6.2 \times
10^{-7} J_0$. The net current integrated over the loop is small,
but non-zero.}
%\centerline{Fig. 6}
\end{figure}

%\newpage 

\begin{figure}
\centering
\includegraphics{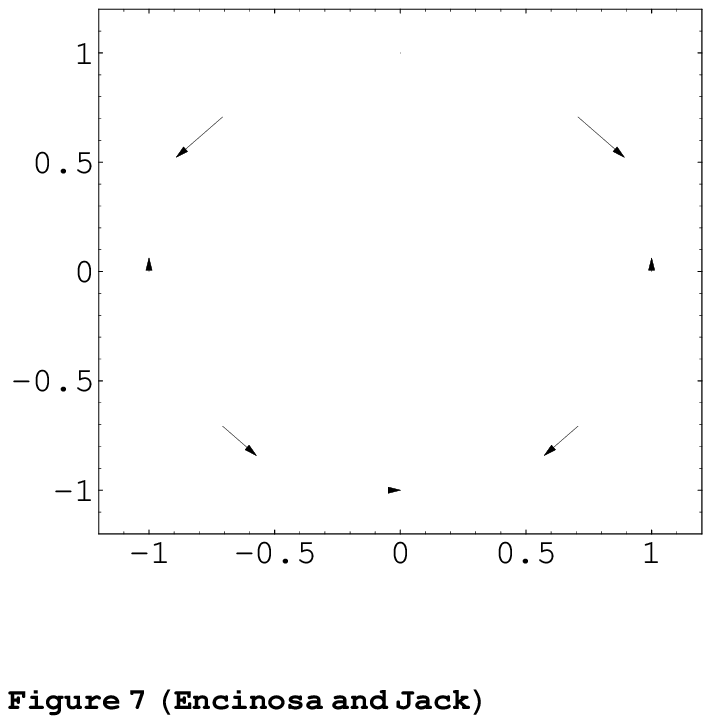}
\caption{$J_\phi(t_f,\phi)$ plotted on the
torus (in units of $R=1$) at $\theta = 0$ in $\phi = \pi/2$
intervals for the LPW case. The longest arrow is $.011 J_0$. The
net current integrated over the loop is zero.}
%\centerline{Fig. 7}
\end{figure}
\begin{figure}
\centering
\includegraphics{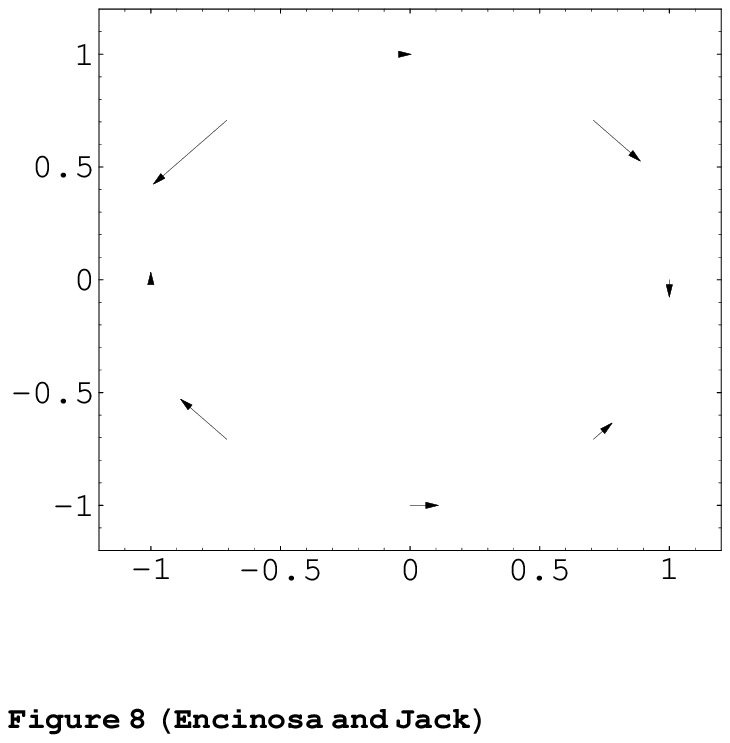}
\caption{$J_\phi(t_f,\phi)$ plotted on the
torus (in units of $R=1$) at $\theta = 0$ in $\phi = \pi/2$
intervals for the CPW case. The net current integrated over the
loop is non-zero. The longest arrow is $.11 J_0$.}
%\centerline{Fig. 8}
\end{figure}

%\newpage

\begin{figure}
\centering
\includegraphics{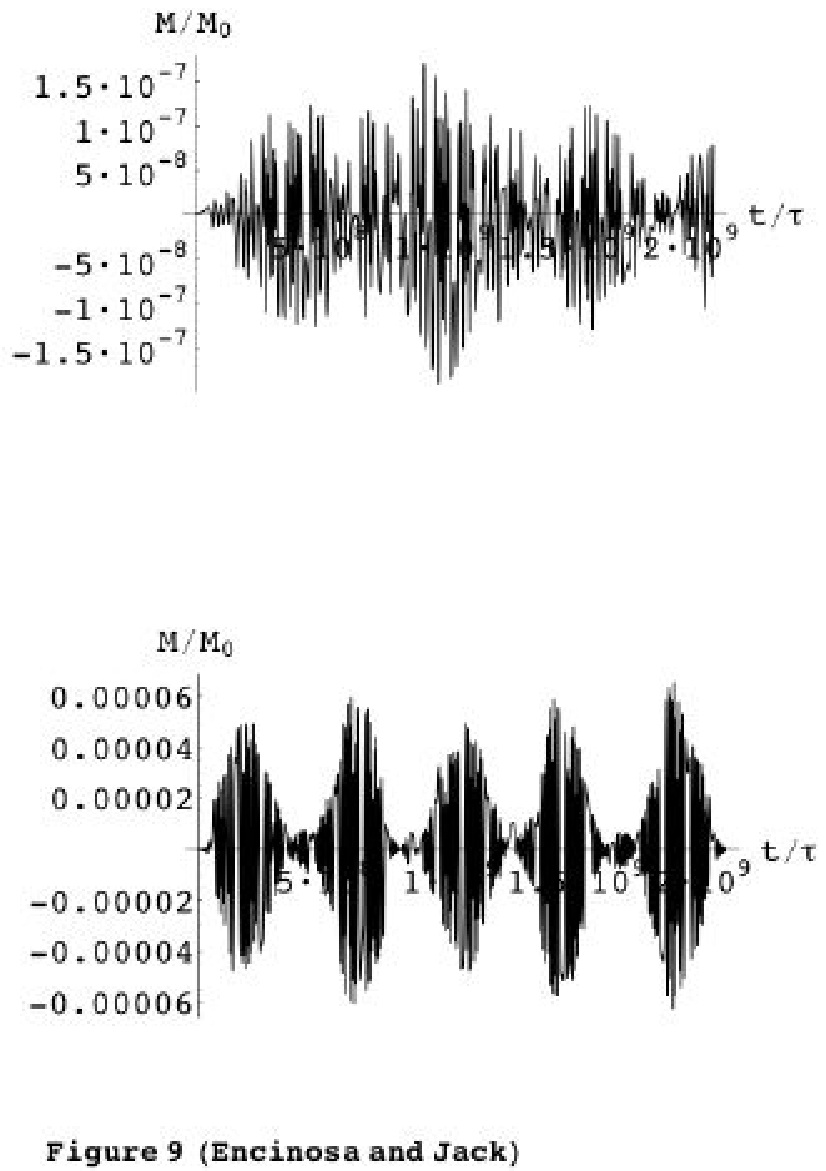}
\caption{Magnetic moments $M_z(t)$ (top)
and $M_\phi(t)$ (lower) in units of $ M_0 = e \hbar R^2 / 4m $ at
$\phi = \pi/2$ for the LPW case. The longest arrow is $.009 J_0$.}
%\centerline{Fig. 9}
\end{figure}
\begin{figure}
\centering
\includegraphics{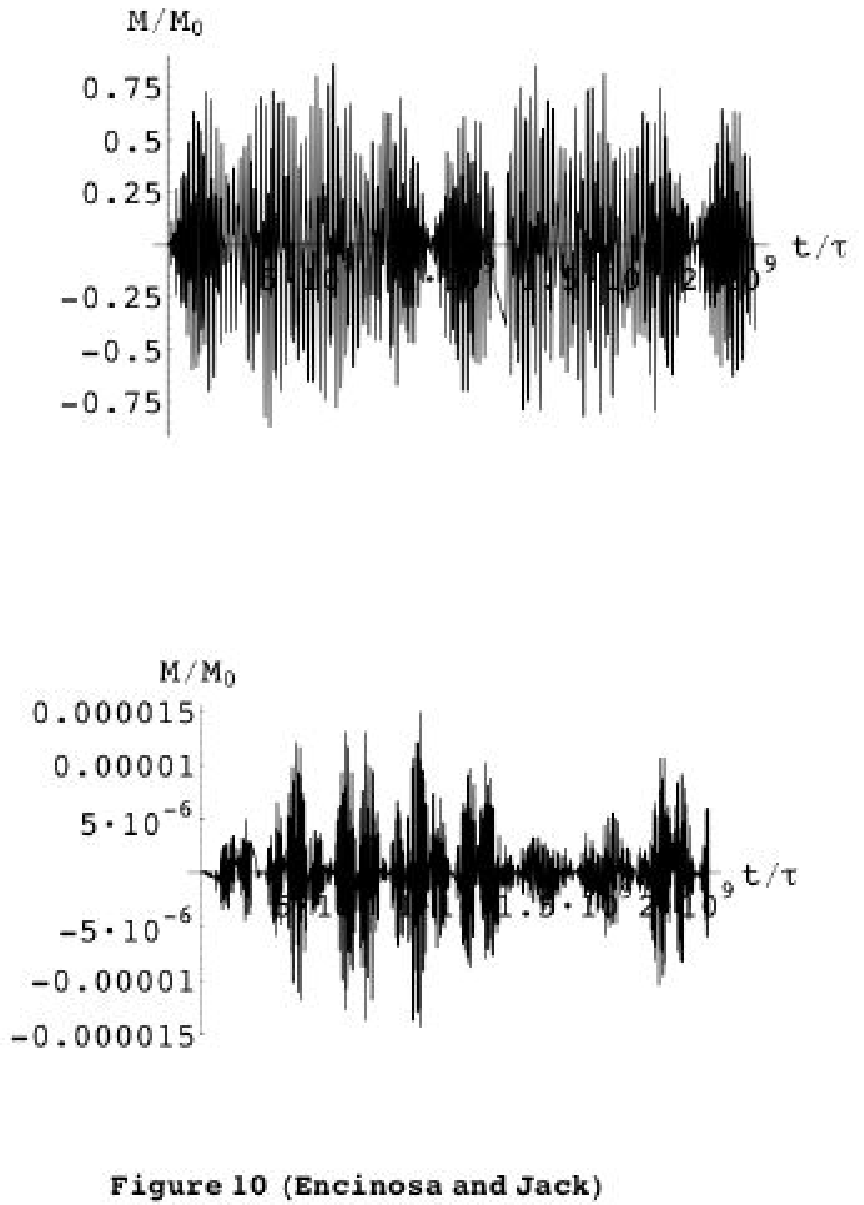}
\caption{Magnetic moments $M_z(t)$ (top)
and $M_\phi(t)$ (lower) in units of $ M_0 = e \hbar R^2 / 4m $ at
$\phi = \pi/2$ for the CPW case.}
%\centerline{Fig. 10}
\end{figure}

%\newpage

\begin{figure}
\centering
\includegraphics{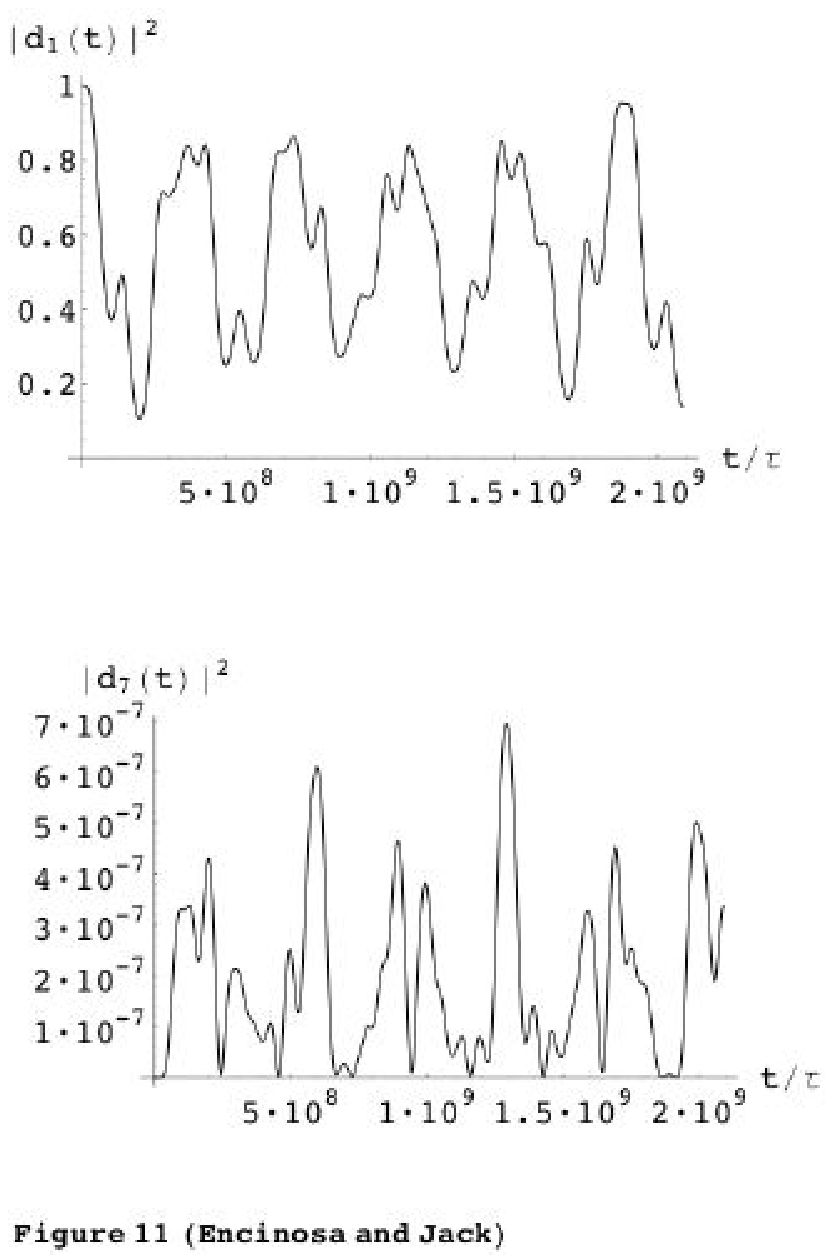}
\caption{Ground state $|d_1(t)|^2$ and
seventh state $|d_7(t)|^2$ time dependent probability amplitudes
for the system subjected to a LPW allowed to interfere with its
reflection off an ideal mirror.}
%\centerline{Fig. 11}
\end{figure}
\begin{figure}
\centering
\includegraphics{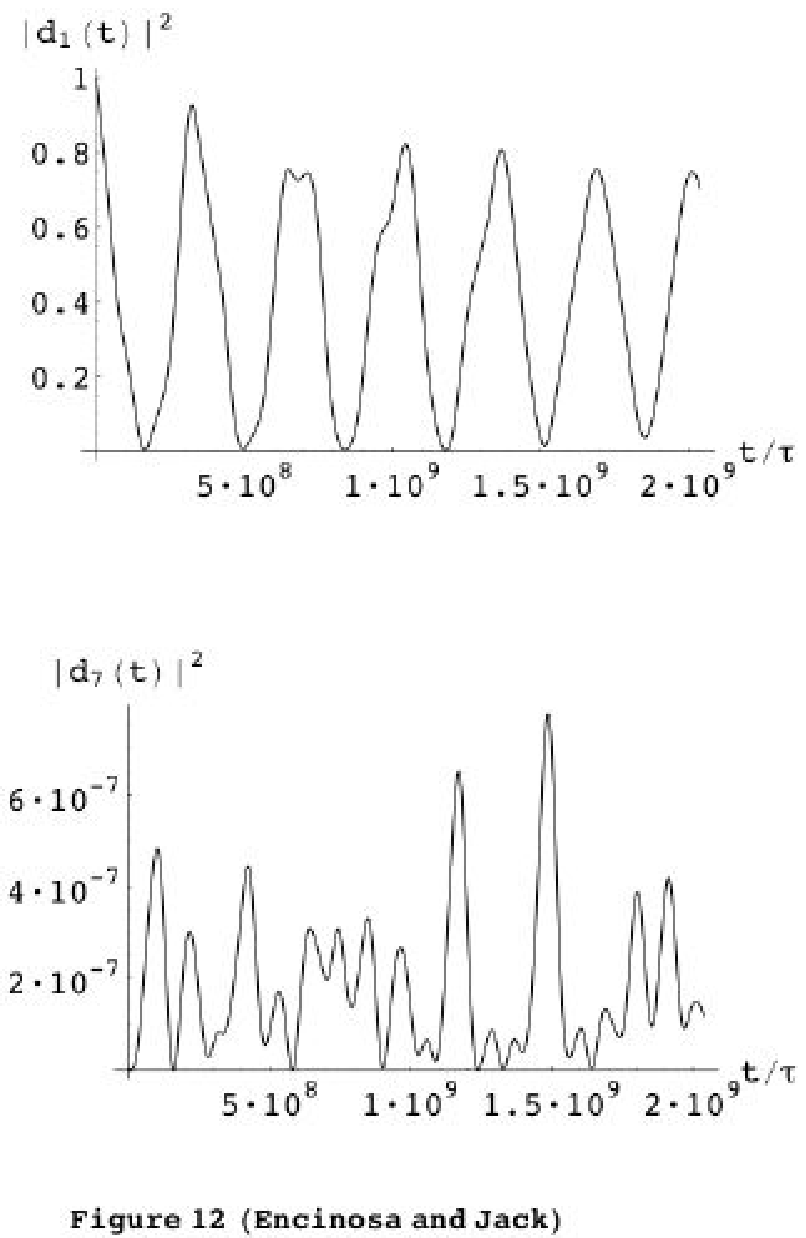}
\caption{Ground state $|d_1(t)|^2$ and
seventh state $|d_7(t)|^2$ time dependent probability amplitudes
for the  system subjected to a CPW allowed to interfere with its
reflection off an ideal mirror.}
%\centerline{Fig. 12}
\end{figure}

%\newpage

\begin{figure}
\centering
\includegraphics{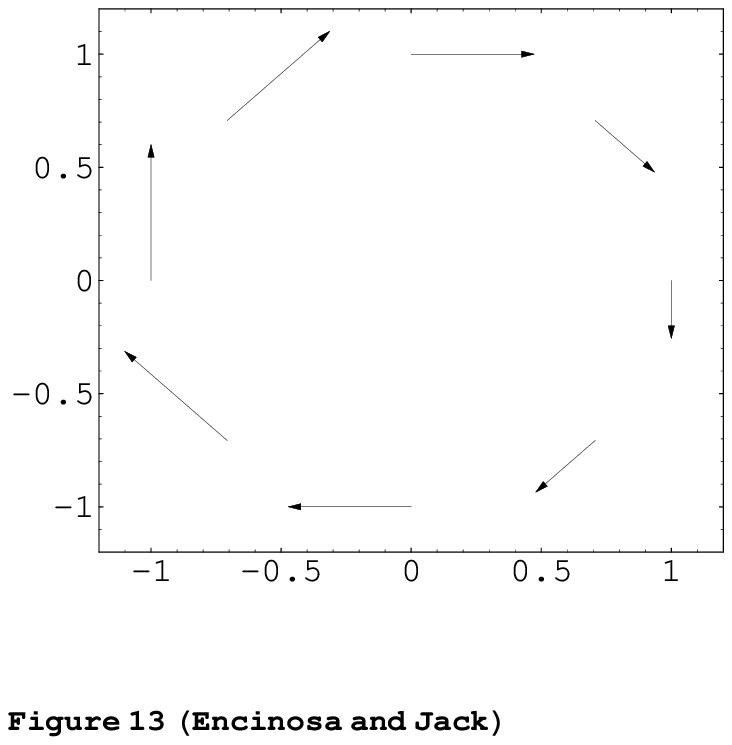}
\caption{$J^b_\theta(t_f,\theta)$ plotted
on the torus (in units of $R=1$) at $\phi = 0$ in $\theta = \pi/4$
intervals for the CPW case where the wave is allowed to interfere
with its reflection. A circulating SCD results. The longest arrow
is $1.2 \times 10^{-6} J_0$.}
%\centerline{Fig. 13}
\end{figure}
\begin{figure}
\centering
\includegraphics{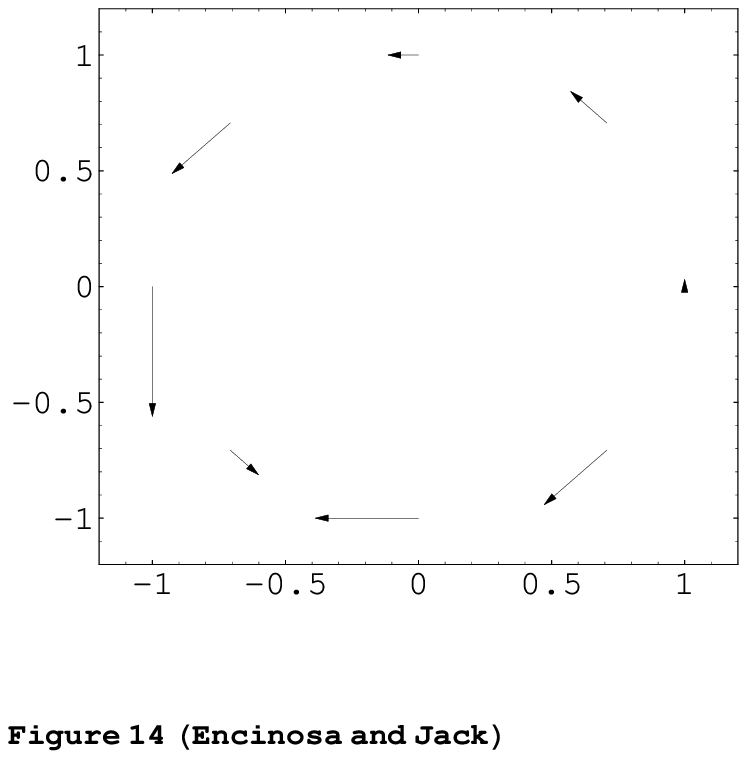}
\caption{$J^b_\phi(t_f,\phi)$ plotted on
the torus (in units of $R=1$) at $\theta = \pi$ in $\phi = \pi/4$
intervals for the CPW case where the wave is allowed to interfere
with its reflection, showing that a circulating SCD results in the
azimuthal direction as well as around the minor radius as per Fig.
13. The longest arrow is $1.5 \times 10^{-2} J_0$.}
%\centerline{Fig. 14}
\end{figure}

%\newpage 

\begin{figure}
\centering
\includegraphics{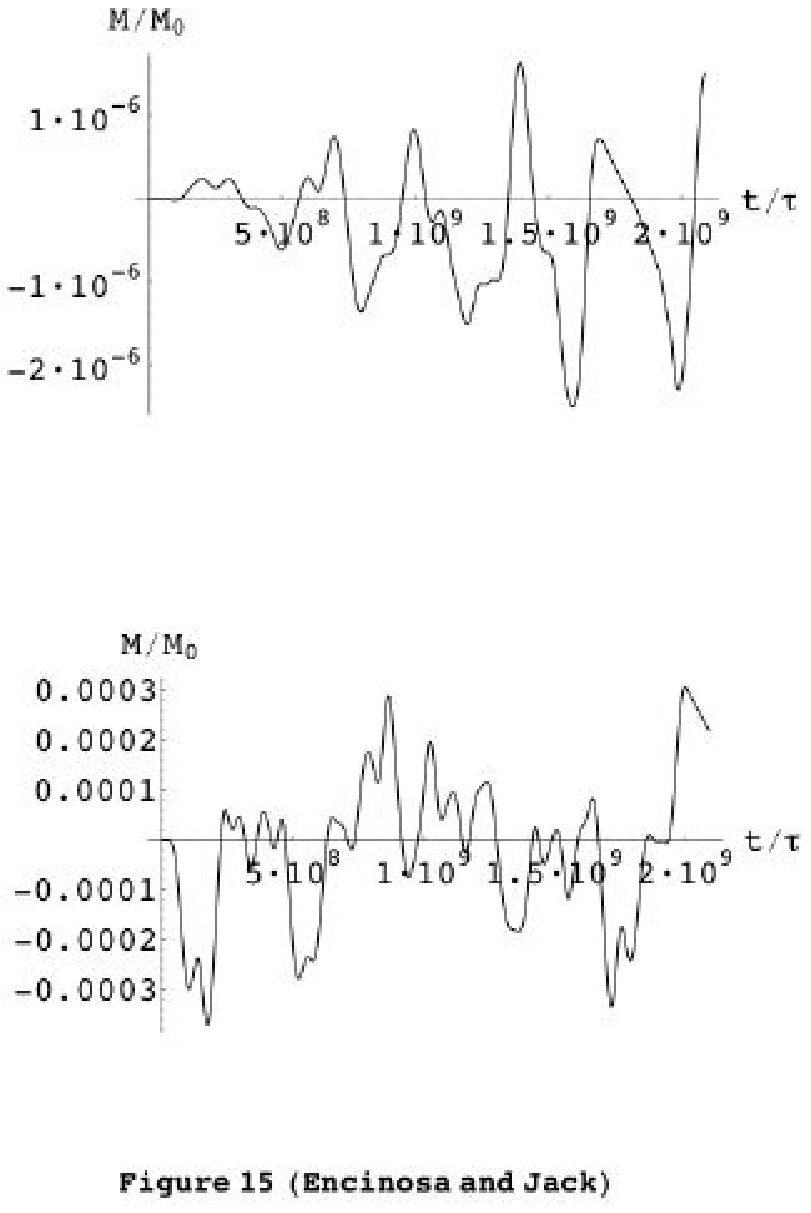}
\caption{Magnetic moments $M_z(t)$ (top)
and $M_\phi(t)$ (lower) in units of $ M_0 = e \hbar R^2 / 4m $ at
$\phi = \pi/2$ for the LPW case where the wave is allowed to
interfere with its reflection.}
%\centerline{Fig. 15}
\end{figure}
\begin{figure}
\centering
\includegraphics{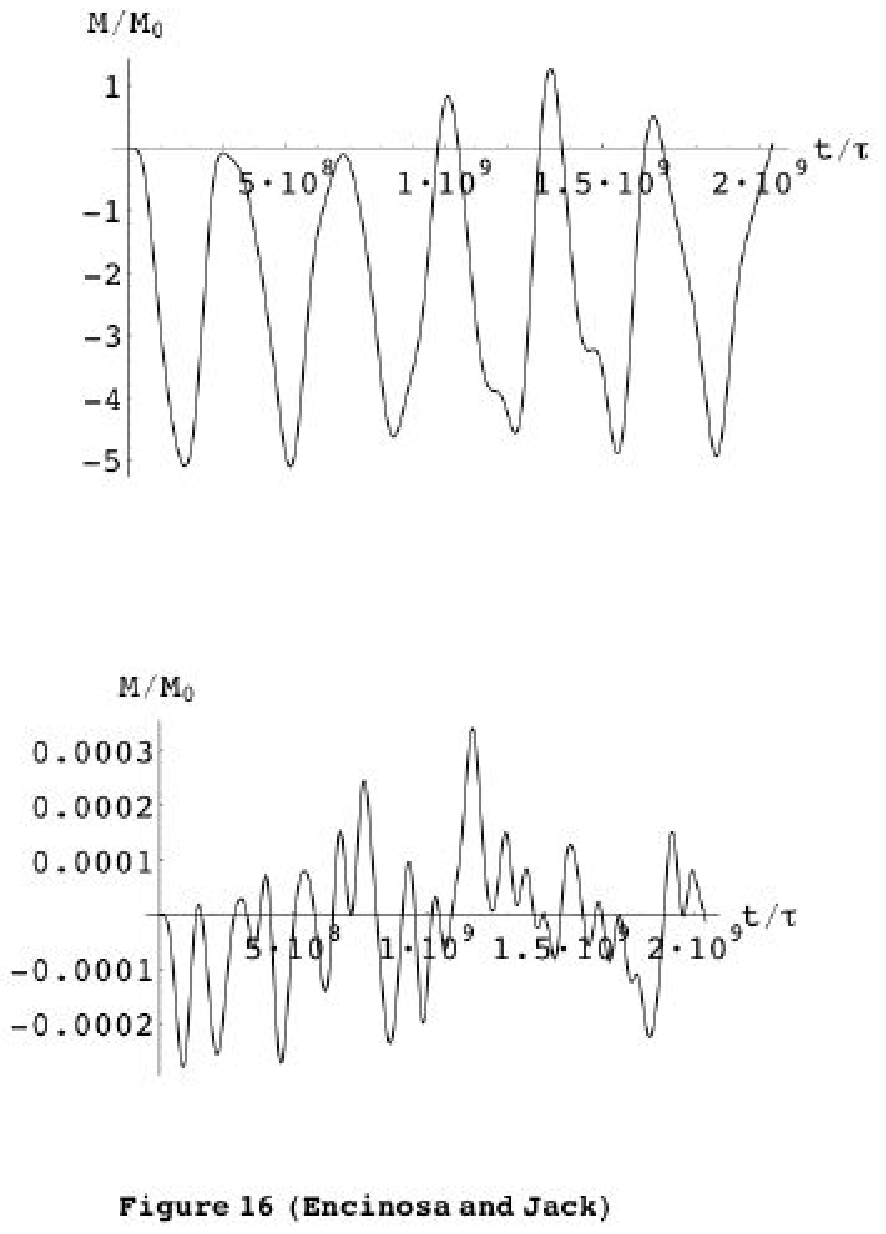}
\caption{Magnetic moments $M_z(t)$ (top)
and $M_\phi(t)$ (lower) in units of $ M_0 = e \hbar R^2 / 4m $ at
$\phi = \pi/2$ for the CPW case where the wave is allowed to
interfere with its reflection.}
%\centerline{Fig. 16}
\end{figure}

\end{document}